\documentclass[pra,twocolumn,superscriptaddress,amsmath,showpacs]{revtex4}
\usepackage{dsfont}
\usepackage{graphicx}
\usepackage{subfigure}

\newcommand{\ketbra}[2]{| #1 \rangle \langle #2 |}

\begin{document}
\title{Gaussian states under coarse-grained continuous variable measurements}
\author{Jiyong Park}
\affiliation{Department of Physics, Texas A\&M University at Qatar, Education City, P.O.Box 23874, Doha, Qatar}
\author{Se-Wan Ji}
\affiliation{Department of Physics, Texas A\&M University at Qatar, Education City, P.O.Box 23874, Doha, Qatar}
\author{Jaehak Lee}
\affiliation{Department of Physics, Texas A\&M University at Qatar, Education City, P.O.Box 23874, Doha, Qatar}
\author{Hyunchul Nha}
\affiliation{Department of Physics, Texas A\&M University at Qatar, Education City, P.O.Box 23874, Doha, Qatar}
\affiliation{School of Computational Sciences, Korea Institute for Advanced Study, Seoul 130-722, Korea}

\begin{abstract}
The quantum-to-classical transition of a quantum state is a topic of great interest in fundamental and practical aspects. A coarse-graining in quantum measurement has recently been suggested as its possible account in addition to the usual decoherence model. 
We here investigate the reconstruction of a Gaussian state (single mode and two modes) by coarse-grained homodyne measurements. To this aim, we employ two methods, the direct reconstruction of the covariance matrix and the maximum likelihood estimation (MLE), respectively, and examine the reconstructed state under each scheme compared to the state interacting with a Gaussian (squeezed thermal) reservoir. We clearly demonstrate that the coarse-graining model, though applied equally to all quadrature amplitudes, is not compatible with the decoherence model by a thermal (phase-insensitive) reservoir. Furthermore, we compare the performance of the direct reconstruction and the MLE methods by investigating the fidelity and the nonclassicality of the reconstructed states and show that the MLE method can generally yield a more reliable reconstruction, particularly without information on a reference frame (phase of input state).
\end{abstract}

\maketitle

\section{introduction}
The discrepancy between quantum and classical mechanics over the description of physical phenomena  has long been an object of interest and controversy. Although quantum mechanics has been successful in describing and manipulating a microscopic world, a macroscopic world can interestingly be explained by classical mechanics that has different premises and framework from quantum mechanics. There has thus been much interest in accounting for the quantum-to-classical transition, and in particular, the decoherence by environmental interactions is nowadays perceived as one of the most promising models in this respect \cite{Zurek1991, Zurek2003}. 

Recently, there have also been some different attempts to explain the quantum-to-classical transition \cite{Kofler2007, Raeisi2011, Wang2013, Jeong2014}. In contrast to the decoherence program, these focus on the inefficiency of quantum measurement, namely, coarse-grained outcomes by imperfect detectors \cite{Kofler2007, Raeisi2011} or imprecise control of target operations \cite{Wang2013, Jeong2014}. Comparing these approaches to the usual decoherence model is thus important to extending our understanding of the quantum-to-classical transition.

In this paper, we investigate single-mode and two-mode Gaussian states under the coarse-graining in the homodyne measurement. 
Gaussian states and operations provide crucial elements of quantum information processing for continuous variables and have been extensively studied both theoretically and experimentally \cite{Weedbrook2012}. Our coarse-graining model is similar to the Ehrenfest's idea of coarse graining \cite{Ehrenfest1912, Ehrenfest1990}, and recently the same model has been considered in the context of the uncertainty relation \cite{Rudnicki2012A, Rudnicki2012B} and the entanglement detection \cite{Tasca2013}. Unlike the last of these \cite{Rudnicki2012A, Rudnicki2012B, Tasca2013}, where the obtained data do not fully characterize the state under investigation, we are interested in quantum state tomography: the process of inferring the prepared quantum state from the measured data \cite{Lvovsky2009}. Reconstructing the density matrix or the phase-space distribution of a quantum state, the process endeavors to provide the maximal information about the given state, which can also be used to verify nonclassical features, e.g. negativity in phase-space and entanglement. Using the coarse-grained data from homodyne detection, we may reconstruct a Gaussian state and compare it to the same state under a Gaussian noisy channel (squeezed thermal environment), thereby comparing the coarse-graining model and the decoherence model in view of the quantum-to-classical transition.

In quantum optics, the inverse Radon transformation of the marginal distribution acquired from homodyne detection was theoretically proposed \cite{Vogel1989} and experimentally implemented \cite{Smithey1993, Beck1993} to reconstruct the Wigner distribution of a given state. However, the direct application of the inverse Radon transformation yields an unphysical state due to the unavoidable process of data binning \cite{D'Ariano1994}. To assure the legitimacy of the reconstructed state, quantum state estimation, which is to determine the most probable physical state from the measured data, was proposed \cite{Hradil1997} and has been employed in experiments \cite{Banaszek1999, Zavatta2004, Babichev2004, Ourjoumtsev2006}. We here employ two methods for state reconstruction under coarse-graining, namely, a direct reconstruction of the covariance matrix and a maximum likelihood estimation (MLE) \cite{Hradil1997}. The coarse graining is equally applied to the homodyne measurement of each quadrature amplitude, and is therefore isotropic in phase space.  One might then expect that there can exist an equivalent decoherence model by a thermal reservoir, more precisely, a phase-insensitive Gaussian reservoir. We, however, show that it is not the case.

Furthermore, we investigate the performance of two reconstruction methods by examining the fidelity between an input state and the reconstructed state and the nonclassicality (squeezing or entanglement) of the reconstructed state. In a realistic situation, sharing the reference frames between the preparer and the verifier can be a critical issue. We thus study how this issue can particularly affect the performance of the direct reconstruction method by considering cases with and without information on the phase of the input state. 

\section{preliminaries}
To begin with, we first introduce our coarse-graining model with homodyne mesurements and the decoherence model with an environmental interaction, respectively.
\subsection{Homodyne measurement under coarse-graining}
A homodyne detector measures the quadrature amplitude $\hat{X}_{\varphi} = ( \hat{a}^{\dag} e^{i \varphi} + \hat{a} e^{- i \varphi} ) / 2$ of an optical field, where $\hat{a}$ ($\hat{a}^\dag$) is the annihilation (creation) operator and $\varphi$ is the phase determined by a local oscillator. The probability distribution $P ( x_\varphi )$ of the amplitude $x_\varphi$ is given by \cite{Vogel1989}
	\begin{equation} \label{eq:characteristic-to-homodyne}
		P ( x_\varphi ) = \frac{1}{\pi} \int_{- \infty}^{\infty} dk C ( \lambda = i k e^{i \varphi} ) e^{- 2 i k x},
		\end{equation}
where $C ( \lambda )$ is the characteristic function of the state $\rho$,
	\begin{equation}
		C ( \lambda ) = \mathrm{tr} [ \rho \hat{D} ( \lambda ) ],
	\end{equation}
with the displacement operator $\hat{D} ( \lambda ) = \exp ( \lambda \hat{a}^{\dag} - \lambda^{*} \hat{a} )$. The characteristic function $C ( \lambda )$ contains the full information on the state $\rho$. In turn, a complete set of homodyne measurements over all phase angles $\varphi \in [ 0, \pi ]$ can be used to construct the density matrix $\rho$ or equivalently its phase-space distributions.

Suppose now that the homodyne measurement does not yield a smooth continuous distribution due to the inefficiency of photodetectors. More precisely, if the measurement cannot distinguish the values of $x_\varphi$ within an interval of size $\sigma$, similar to the Ehrenfest's idea of coarse-graining \cite{Ehrenfest1912, Ehrenfest1990}, we obtain a coarse-grained probability distribution as
	\begin{equation} \label{eq:coarse-graining}
		P_D( x_\varphi ) = \sum_{m = - \infty}^{\infty} P_{\sigma} [ m, \varphi ] \mathrm{rect} \bigg( \frac{x}{\sigma} - m \bigg).
	\end{equation}
Here, $\mathrm{rect} ( x )$ is a step function
	\begin{equation}
		\mathrm{rect} ( x ) = \left\{
			\begin{array}{ll}
				0 & \text{for } |x| > 1/2, \\
				1 & \text{for } |x| \leq 1/2,
			\end{array}
		\right.
	\end{equation}
and $P_{\sigma} [ m, \varphi ]$ represents the coarse-grained (averaged) probability in the region of $x\in [(m-\frac{1}{2})\sigma,(m+\frac{1}{2})\sigma]$  as
	\begin{equation}
		P_{\sigma} [ m, \varphi ]\equiv \frac{1}{\sigma}\int_{( m - \frac{1}{2} ) \sigma}^{( m + \frac{1}{2} ) \sigma} dx P ( x_\varphi ),
	\end{equation}
using $P ( x_\varphi )$ from Eq.~(\ref{eq:characteristic-to-homodyne}). As an example, Fig. 1 illustrates how the coarse-graining process transforms an original distribution $P ( x_\varphi )$ to a piecewise flat distribution $P_D( x_\varphi )$.

	\begin{figure}
		\includegraphics[width=\linewidth]{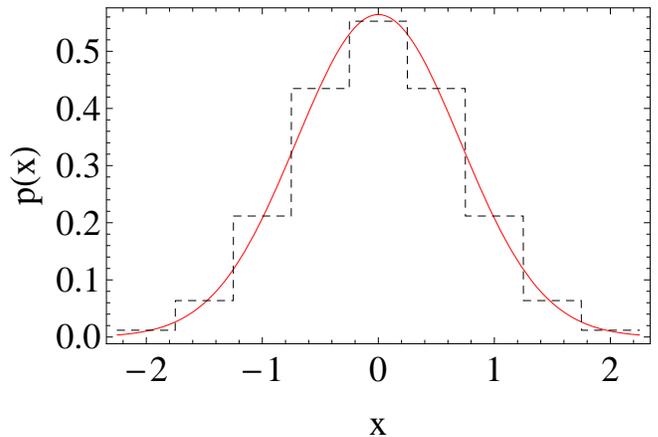}
		\caption{(Color online) Illustration of the coarse-graining process in Eq.~(\ref{eq:coarse-graining}). A Gaussian probability distribution (red solid line) is transformed to a piecewisely flat distribution (black dashed line) under the coarse-graining of size $\sigma=0.5$.}
		\label{fig:0-coarse-graining}
	\end{figure}

In general, it is known that this coarse-grained marginal distribution cannot be directly used to reconstruct a density matrix or its phase-space distributions because the output does not correspond to a physical state \cite{D'Ariano1994}. To reconstruct a legitimate quantum state from the coarse-grained homodyne measurement, we thereby employ an MLE method that is designed to find the most probable physical state by maximizing the log-likelihood estimator
	\begin{equation}
		\mathcal{L} = \int d\mu P_{D} \ln P_{E},
	\end{equation}
where $\mu$ is a probability measure, $P_{D}$ the probability distribution obtained from measurement, and $P_{E}$ is the probability distribution from an estimated state. From the perspective of information theory, the method can be seen as the minimization of the relative entropy of two distributions
	\begin{equation}
		D ( P_{D} || P_{E} ) \equiv \int d\mu P_{D} \ln \frac{P_{D}}{P_{E}},
	\end{equation}
that is, we optimize $P_{E}$  for a given $P_{D}$ to obtain a minimal value of $D ( P_{D} || P_{E} )$. The relative entropy becomes zero if and only if $P_{D} = P_{E}$, that is, only when the obtained data can correspond to a certain physical state.

\subsection{$n$-mode Gaussian states under Gaussian reservoirs}
An $n$-mode Gaussian state is fully identified by its first and second moments (for a review, see Ref. \cite{Weedbrook2012}). It has a Gaussian characteristic function in the form
	\begin{align}
		C ( \boldsymbol{\lambda} ) & \equiv \mathrm{tr} [ \rho \prod_{i = 1}^{n} \hat{D}_{i} ( \lambda_{i} ) ] \nonumber \\
		& = \exp ( - \boldsymbol{\lambda} \boldsymbol{\Gamma} \boldsymbol{\lambda}^{T} + i \sqrt{2} \langle \hat{\mathbf{R}} \rangle \boldsymbol{\lambda}^{T} ),
	\end{align}
where $\hat{\mathbf{R}} \equiv ( \hat{q}_{1}, \hat{p}_{1}, ..., \hat{q}_{n}, \hat{p}_{n} )$ is related to the quadrature amplitudes $\hat{X}_{i, 0} = \hat{q_{i}} / \sqrt{2}$ and $\hat{X}_{i, \pi / 2} = \hat{p}_{i} / \sqrt{2}$ with $i \in \{ 1, ..., n \}$, $\boldsymbol{\lambda} = ( \Im [ \lambda_{1} ], - \Re [ \lambda_{1} ], ..., \Im [ \lambda_{n} ], - \Re [ \lambda_{n} ] )$, with $\mathrm{Re} [ \lambda_{i} ]$ and $\mathrm{Im} [ \lambda_{i} ]$ the real part and the imaginary part of $\lambda_{i}$, respectively.  $\boldsymbol{\Gamma}$ is the covariance matrix whose elements are
	\begin{equation}
		\Gamma_{ij} = \frac{1}{2} \langle \hat{R}_{i} \hat{R}_{j} + \hat{R}_{j} \hat{R}_{i} \rangle - \langle \hat{R}_{i} \rangle \langle \hat{R}_{j} \rangle,
	\end{equation}
where $\langle \hat{o} \rangle \equiv \mathrm{tr} ( \rho \hat{o} )$ is the expectation value of the operator $\hat{o}$.

A Gaussian process transforms a Gaussian state into another Gaussian state, and a typical Gaussian process is the environmental interaction with Gaussian (thermal squeezed) reservoirs, which usually leads to decoherence. This decoherence process can be described by a master equation
	\begin{align} \label{eq:master-equation}
		\dot{\rho} ( t ) = \sum_{i = 1}^{n} \frac{\gamma_{i}}{2} \{ ( N_{i} + 1 ) \mathcal{L} [ \hat{a} ] + N_{i} \mathcal{L} [ \hat{a}^{\dag} ] & \nonumber \\
		- M_{i}^{*} \mathcal{D} [ \hat{a} ] - M_{i} D [ \hat{a}^{\dag} ]  & \} \rho ( t ),
	\end{align}
where $\gamma_{i}$ is the interaction strength for the $i$-th mode, and $\mathcal{L} [ \hat{o} ] \rho = 2 \hat{o} \rho \hat{o}^{\dag} - \hat{o}^{\dag} \hat{o} \rho - \rho \hat{o}^{\dag} \hat{o}$ and $\mathcal{D} [ \hat{o} ] \rho = 2 \hat{o} \rho \hat{o} - \hat{o} \hat{o} \rho - \rho \hat{o} \hat{o}$ are Lindblad superoperators. The covariance matrix of the reservoir interacting with the $i$-th mode is given by
	\begin{equation}
		\boldsymbol{\Gamma}_{i, r} =
		\begin{pmatrix}
			\frac{1}{2} + N_{i} + \Re [ M_{i} ] & \Im [ M_{i} ] \\
			\Im [ M_{i} ] & \frac{1}{2} + N_{i} - \Re [ M_{i} ]
		\end{pmatrix},
	\end{equation}
with $N_{i}$ and $M_{i}$ representing the mean thermal photon number and the squeezing parameter of the reservoir, respectively. The master equation in Eq.~\eqref{eq:master-equation} can be converted into a differential equation for the characteristic function
	\begin{equation} \label{eq:differential-equation}
		\frac{\partial}{\partial t} C ( \boldsymbol{\lambda}, t ) = - \sum_{i = 1}^{n} \frac{\gamma_{i}}{2} ( A_{i} + B_{i} ) C ( \boldsymbol{\lambda}, t ).
	\end{equation}
where
	\begin{align}
		A_{i} & = ( 1 + 2 N_{i} ) | \lambda_{i} |^{2} - M_{i} ( \lambda_{i}^{*} )^{2} - M_{i}^{*} \lambda_{i}^{2}, \nonumber \\
		B_{i} & = \lambda_{i}^{*} \frac{\partial}{\partial \lambda_{i}^{*}} + \lambda_{i} \frac{\partial}{\partial \lambda_{i}}.
	\end{align}
The solution to Eq.~\eqref{eq:differential-equation} can be represented in terms of the covariance matrix $\Gamma_{\rho} ( t )$ of the state at time $t$
	\begin{equation} \label{eq:time-evolution-of-covariance-matrix}
		\boldsymbol{\Gamma}_{\rho} ( t ) = \sqrt{\mathbf{G}} [ \boldsymbol{\Gamma}_{\rho} ( 0 ) - \mathbf{\Gamma}_{r} ] \sqrt{\mathbf{G}} + \mathbf{\Gamma}_{r},
	\end{equation}
where $\mathbf{G} = \bigoplus_{i = 1}^{n} \exp ( - \gamma_{i} t ) \mathds{1}_{2}$ and $\boldsymbol{\Gamma}_{r} = \bigoplus_{i = 1}^{n} \boldsymbol{\Gamma}_{i, r}$.

\section{single-mode Gaussian state estimation}
In this section, we investigate single-mode Gaussian states reconstructed from the coarse-grained homodyne data. In general, as mentioned before, one can reconstruct a given state by measuring the probability distributions of quadrature amplitudes for all (practically speaking, many) phase angles and then relying on the Radon transformation \cite{Vogel1989}. We adopt this approach under the coarse grained measurement together with the MLE method. 

\subsection{Direct reconstruction of covariance matrix}
On the other hand, since a Gaussian state is completely identified by its first and second moments, one can also reconstruct the given state by determining only those moments, which will be another approach, namely, a direct reconstruction of the covariance matrix. For the case of ideal homodyne detection, the moments can be determined by measuring only three different quadratures $\hat{X}_{\varphi = 0}$, $\hat{X}_{\varphi = \pi / 4}$, and $\hat{X}_{\varphi = \pi / 2}$ as
	\begin{align} \label{eq:single-mode-direct-reconstruction}
		\Gamma_{11} = & \langle \hat{q}^{2} \rangle - \langle \hat{q} \rangle^{2} = 2 \langle \hat{X}_{\varphi = 0}^{2} \rangle - 2 \langle \hat{X}_{\varphi = 0} \rangle^{2} , \nonumber \\
		\Gamma_{22} = & \langle \hat{p}^{2} \rangle - \langle \hat{p} \rangle^{2} = 2 \langle \hat{X}_{\varphi = \pi / 2}^{2} \rangle - 2 \langle \hat{X}_{\varphi = \pi / 2} \rangle^{2}, \nonumber \\
		\Gamma_{12} = & \frac{1}{2} \langle \hat{q} \hat{p} + \hat{p} \hat{q} \rangle - \langle \hat{q} \rangle \langle \hat{p} \rangle \nonumber \\
		= & 2 \langle \hat{X}_{\varphi = \pi / 4}^{2} \rangle - \langle \hat{X}_{\varphi = 0}^{2} \rangle - \langle \hat{X}_{\varphi = \pi / 2}^{2} \rangle \nonumber \\
		- & 2 \langle \hat{X}_{\varphi = 0} \rangle \langle \hat{X}_{\varphi = \pi / 2} \rangle,
	\end{align}
where the $n$-th moment of the quadrature $\hat{X}_{\varphi}$ is given by $\langle \hat{X}^{n}_{\varphi} \rangle = \int dx_{\varphi} x_{\varphi}^{n} P ( x_{\varphi} )$ with a relevant probability distribution $P ( x_{\varphi} )$.	
	
An arbitrary single-mode Gaussian state can be expressed as a displaced squeezed thermal state in the form
	\begin{equation}
		\rho = \hat{D} ( \alpha ) \hat{S} ( r, \phi_i ) \rho_{th} ( \bar{n} ) \hat{S}^{\dag} ( r, \phi_i ) \hat{D}^{\dag} ( \alpha).
	\end{equation}
Here $\hat{S} ( r, \phi_i ) = \exp [ - \frac{r}{2} \{ \exp ( 2i \phi_i ) ( \hat{a}^{\dag} )^{2} - \exp ( - 2i \phi_i ) \hat{a}^{2} \} ]$ is the squeezing operator with the squeezing strength $r$, the angle $\phi_i$ of the squeezing axis, and $\rho_{th} ( \bar{n} )$ is the thermal state with the mean photon number $\bar{n}$:
	\begin{equation}
		\rho_{th} ( \bar{n} ) = \sum_{n = 0}^{\infty} \frac{\bar{n}^{n}}{( \bar{n} + 1 )^{n + 1}} \ketbra{n}{n}.
	\end{equation}

For a squeezed thermal state, the covariance matrix is given by
	\begin{align} \label{eq:displaced-squeezed-thermal-state}
		\Gamma_{11} & = ( \bar{n} + \frac{1}{2} ) [ \cosh ( 2r ) - \sinh ( 2r ) \cos 2\phi_i ], \nonumber \\
		\Gamma_{22} & = ( \bar{n} + \frac{1}{2} ) [ \cosh ( 2r ) + \sinh ( 2r ) \cos 2\phi_i ], \nonumber \\
		\Gamma_{12} & = - ( \bar{n} + \frac{1}{2} ) \sinh ( 2r ) \sin 2\phi_i,
	\end{align}
and its characteristic function can be expressed as
	\begin{equation}
		C_{1} ( \lambda ) = \exp ( - \Gamma_{22} \lambda_{r}^{2} - \Gamma_{11} \lambda_{i}^{2} + 2 \Gamma_{12} \lambda_{r} \lambda_{i} ),
	\end{equation}
with $\lambda_{r}=\Re[\lambda]$ and $\lambda_{i}=\Im[\lambda]$. The corresponding homodyne distribution is then given by
	\begin{eqnarray}
		P ( x_\varphi ) &=& \sqrt{\frac{1}{2\pi \Delta^2}} \exp \bigg( - \frac{x_\varphi^{2}}{2\Delta^2} \bigg),\nonumber\\
		2\Delta^2 &=& ( \bar{n} + \frac{1}{2} ) [ \cosh ( 2r ) - \sinh ( 2r ) \cos ( 2 \varphi - 2\phi_i ) ].\nonumber\\
	\end{eqnarray}
Inverting relations in Eq.~\eqref{eq:displaced-squeezed-thermal-state}, we obtain
	\begin{align} \label{eq:single-mode-parameters}
		\bar{n} & = \sqrt{\det \boldsymbol{\Gamma}} - \frac{1}{2}, \nonumber \\
		r & = \frac{1}{2} \mathrm{arcsinh} \Bigg( \frac{1}{2} \sqrt{\frac{\gamma}{\det \boldsymbol{\Gamma}}} \Bigg), \nonumber \\
		2\phi_i & = \begin{cases} - \arcsin \Bigg( \dfrac{2 \Gamma_{12}}{\sqrt{\gamma}} \Bigg) & \textrm{for} \quad \Gamma_{11} \leq \Gamma_{22}, \\
		\pi + \arcsin \Bigg( \dfrac{2 \Gamma_{12}}{\sqrt{\gamma}} \Bigg) & \textrm{for} \quad \Gamma_{11} > \Gamma_{22}, \end{cases}
	\end{align}
with $\det \boldsymbol{\Gamma} = \Gamma_{11} \Gamma_{22} - \Gamma_{12}^{2}$ and $\gamma = ( \Gamma_{22} - \Gamma_{11} )^{2} + 4 \Gamma_{12}^{2}$. 

Using Eq.~\eqref{eq:single-mode-parameters}, we can determine the parameters $(\bar{n},r,\phi_i)$ characterizing a single-mode Gaussian state [Eq. (16)] by simply measuring three quadratures in Eq.~\eqref{eq:single-mode-direct-reconstruction}, which will be used under coarse-grained measurements.
	\begin{figure}
		\includegraphics[width=\linewidth]{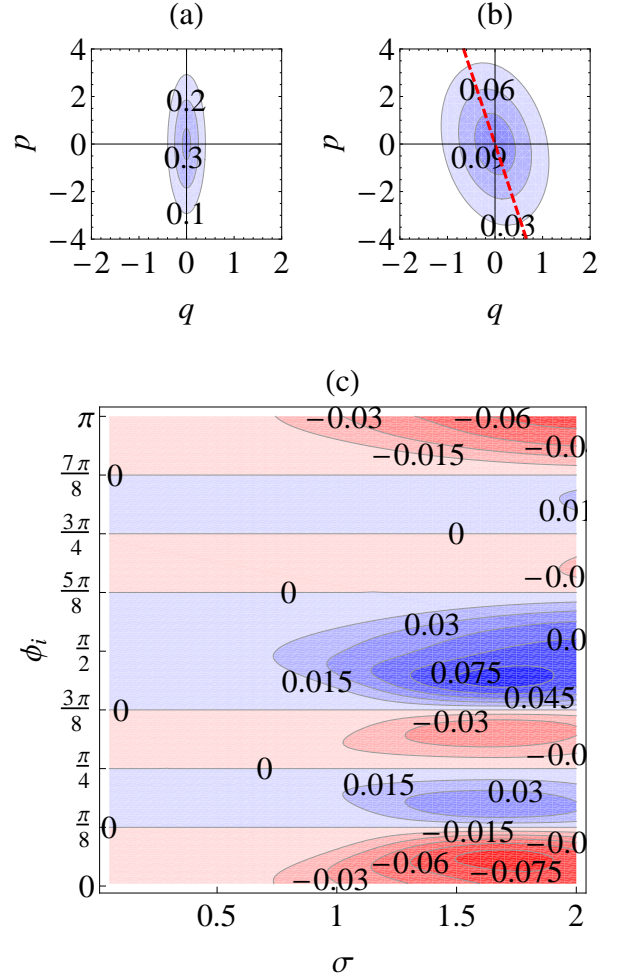}
		\caption{(Color online) (a) Original squeezed state (b) reconstructed state with coarse-graining $\sigma=0.1$ (c) difference in squeezing angle between an input state and its estimated state as a function of the coarse-graining size $\sigma$ and the input squeezing angle $\phi_i$. In all plots, the input state has the parameters $( \bar{n}, r ) = ( 0, 1 )$.}
		\label{fig:1-direct-reconstruction}
	\end{figure}
From now on, we assume that our measurement settings are fixed to measure three quadratures corresponding to the angles $\varphi=0,\pi/4,\pi/2$ in Eq. (15), whereas the input squeezing angle $\phi_i$ in Eq. (16) is unknown to an experimenter.
	
Note that the finite precision $\sigma$ of homodyne measurement under coarse-graining induces noise to the various moments in Eq. (15), thereby degrading the information on the elements $\Gamma_{ij}$ of the covariance matrix. 
Specifically, using Eqs. (3) to (5), the variance of the measured quadrature can be decomposed into 
\begin{equation}
		\Delta^2 X_\sigma = \frac{\sigma^{2}}{12} + \Delta^2 X_m 
	\end{equation}
where the first term $\frac{\sigma^{2}}{12}$ represents the variance of a flat distribution of size $\sigma$ and the second term the variance of the discretized probability distribution centered at each value $x_m\equiv m\sigma$ \cite{Rudnicki2012A, Rudnicki2012B}. That is,
\begin{equation}
\Delta^2 X_m\equiv \sum_{m = - \infty}^{\infty} x_m ^{2} P(x_m)-\left(\sum_{m = - \infty}^{\infty} x_m P(x_m)\right)^2,
\end{equation}
where the discrete distribution is given by $P(x_m)=\sigma P_\sigma[m]$ using Eq. (5).
For the case of the initial Gaussian distribution with variance $V(0)$, we have 
\begin{align}
		& P_{\sigma} [m] \nonumber \\
		& = \frac{1}{2 \sigma} \Big\{ \mathrm{erf} \Big[ \Big( m + \frac{1}{2} \Big) \frac{\sigma_{N}}{\sqrt{2}} \Big] - \mathrm{erf} \Big[ \Big( m - \frac{1}{2} \Big) \frac{\sigma_{N}}{\sqrt{2}} \Big] \Big\},
	\end{align}
with the normalized coarse-graining size $\sigma_{N} \equiv \sigma / \sqrt{V ( 0 )}$.
Under the coarse-grained homodyne detection, therefore, the characterization of the output state using the direct reconstruction method is affected by way of Eq. (22) in conjunction with Eqs. (15) and (21). From Eq. (20), the variance of coherent states is given by $\Delta^2=1/4$, therefore, the scale of $\sigma$ in our consideration is such that $\sigma=1$ takes the homodyne data within the range $2\Delta$ of coherent states into a single bin.

First, as an illustration, we plot an original Gaussian (squeezed) state [Fig. 2(a)] and the reconstructed state under coarse-graining [Fig. 2 (b)]. We can clearly see that the degree of squeezing is degraded due to the inevitable noise introduced by the coarse-grained data. Furthermore, we also see that the squeezing axis is slightly rotated as a result of the coarse-grained construction through Eq.~\eqref{eq:single-mode-direct-reconstruction}.

In Fig.~\ref{fig:1-direct-reconstruction} (c), we plot the difference in the squeezing angle between an input state and its estimated state under coarse-graining 
as a function of the coarse-graining size $\sigma$ and the input squeezing angle $\phi_i$. We have used input squeezed thermal states with $( \bar{n}, r ) = ( 0, 1 )$. The figure shows that the squeezing angle rotates under the estimation process and that the rotation is non-uniform even with the input states of identical $( \bar{n}, r )$. This implies that the information on the reference frame (squeezing direction) of an input state is important in estimating the given state. (See also the plots in Figs. 4 and 5.) 

Note that the rotation of the squeezing axis does not occur for the input squeezing angles 
$\phi_i=\frac{k\pi}{8}$ ($k=1,2,3,5,6,7$) regardless of $\sigma$ [Fig. 2 (c)]. This can be explained by looking into Eq. (21), where the angle $\phi_i$ is determined by the ratio $|(\Gamma_{22}-\Gamma_{11})/2\Gamma_{12}|$.\\ (i) For $\phi_i=\frac{\pi}{4}$ or $\frac{3\pi}{4}$, i.e., the input squeezing is along the direction half way between the $q$ and $p$ axes, we obviously obtain $\Delta^2 \hat{X}_{\varphi = 0} = \Delta^2 \hat{X}_{\varphi = \pi / 2}$ from homodyne measurements. This leads to $\Gamma_{22}-\Gamma_{11}=0$ regardless of $\sigma$. \\ (ii)  For $\phi_i=\frac{\pi}{8}$ or $\frac{5\pi}{8}$, we obtain $\Delta^2 \hat{X}_{\varphi = 0} = \Delta^2 \hat{X}_{\varphi = \pi / 4}$ using Eq. (20), which must be true even with the coarse-graining of the homodyne data. Then, from Eq. (15), we have $2\Gamma_{12}=2\Delta^2 \hat{X}_{\varphi = 0}-2\Delta^2 \hat{X}_{\varphi =\pi / 2}=\Gamma_{11}-\Gamma_{22}$. \\(iii)  For $\phi_i=\frac{3\pi}{8}$ or $\frac{7\pi}{8}$, we obtain $\Delta^2 \hat{X}_{\varphi = \pi/2} = \Delta^2 \hat{X}_{\varphi = \pi / 4}$ using Eq. (20). Then, from Eq. (15), we have $2\Gamma_{12}=-2\Delta^2 \hat{X}_{\varphi = 0}+2\Delta^2 \hat{X}_{\varphi =\pi / 2}=-\Gamma_{11}+\Gamma_{22}$. The above relations do not change even with added noises due to coarse-graining, and the ratio $\left|(\Gamma_{22}-\Gamma_{11})/2\Gamma_{12}\right|$ is unchanged.

\subsection{Maximum-likelihood-estimation Method}

Next we compare the reconstructed coarse-grained Gaussian states with the same input states under a Gaussian reservoir to see if there can be correspondence between the two models. The decoherence by a thermal reservoir adds noise isotropically to all quadratures in phase-space, so it does not change the squeezing direction of the input state. We thus immediately see that our coarse-graining model based on direct reconstruction is not compatible with the decoherence model by a thermal reservoir, and to find out an equivalence, we have to look into the case of a phase-sensitive reservoir, i.e., a squeezed thermal reservoir. Mathematically, note that Eq.~\eqref{eq:time-evolution-of-covariance-matrix} can be simplified to a convex sum of two covariance matrices,
	\begin{equation} \label{eq:time-evolution-for-single-mode}
		\boldsymbol{\Gamma} ( t ) = y \boldsymbol{\Gamma} ( 0 ) + ( 1 - y ) \boldsymbol{\Gamma}_{r},
	\end{equation}
where $y = \exp ( - \gamma t )$ ($\in [0,1]$).

Instead of finding the equivalence between the direct reconstruction model and the decoherence model, we further extend the coarse-graining model to the case of measuring a full set of quadrature amplitudes.
Unlike the direct reconstruction based on only three quadrature distributions in Eq.~\eqref{eq:single-mode-direct-reconstruction}, we may avoid some negative features like the state rotation in phase space if we obtain a full set of homodyne data and employ the MLE method. In this case, the MLE works for the optimization of
	\begin{equation}
		\mathcal{L} = \int_{0}^{\pi} d \varphi \int_{- \infty}^{\infty} dx P_{D} ( x_\varphi ) \ln P_{E} ( x_\varphi ),
	\end{equation}
where $P_{D} ( x_\varphi )$ and $P_{E} ( x_\varphi )$ are the coarse-grained homodyne distribution and an estimated one, respectively. If the estimation process and the decoherence model are to be equivalent, there must exist a solution $y \in [ 0, 1 ]$ of Eq.~\eqref{eq:time-evolution-for-single-mode} for each estimated state as
	\begin{equation} \label{eq:condition-for-equivalence}
		y  = \frac{(2 \bar{n}_{e} + 1 ) \sinh (2 r_{e} ) - ( 2 \bar{n}_{r} +1 ) \sinh ( 2 r_{r} )}{(2 \bar{n}_{i} + 1 ) \sinh (2 r_{i} ) - ( 2 \bar{n}_{r} +1 ) \sinh ( 2 r_{r} )},
	\end{equation}
where the subscripts $e$ and $i$ represent the estimated state and the input state, respectively. If the reservoir is isotropic, that is, a thermal reservoir with no squeezing $r_{r} = 0$, Eq.~\eqref{eq:condition-for-equivalence} can be simplified to
	\begin{equation} \label{eq:isotropic-reservoir}
		y =  \frac{(2 \bar{n}_{e} + 1 ) \sinh ( 2 r_{e} )}{(2 \bar{n}_{i} + 1 ) \sinh ( 2 r_{i} )}.
	\end{equation}

	\begin{figure}
		\includegraphics[width=\linewidth]{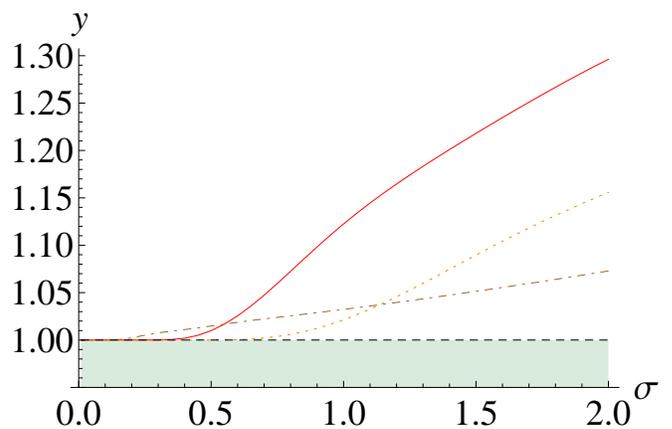}
		\caption{(Color online) The fraction $y$ in Eq.~\eqref{eq:isotropic-reservoir}  to make equal the state estimation process and the decoherence program with an isotropic (thermal) reservoir as a function of coarse-graining size $\sigma$. The input squeezed thermal states are characterized by $( \bar{n}, r ) = ( 0, 1 )$ (red solid line), $( \bar{n}, r ) = ( 1, 1 )$ (orange dotted line), and $( \bar{n}, r ) = ( 0, 2 )$ (brown dot-dashed line).}
		\label{fig:1-likelihood-estimation}
	\end{figure}

In Fig.~\ref{fig:1-likelihood-estimation}, we plot the value of $y$ in Eq.~\eqref{eq:isotropic-reservoir} as a function of coarse-graining size $\sigma$ for the input squeezed thermal states with the parameters $( \bar{n}, r ) = ( 0, 1 )$, $( \bar{n}, r ) = ( 1, 1 )$, and $( \bar{n}, r ) = ( 0, 2 )$. The plot clearly shows $y>1$, therefore, the coarse-graining model cannot be made equivalent to the decoherence model with an isotropic (phase-insensitive) thermal reservoir.

On the other hand, if we consider a squeezed thermal reservoir with $r_{r} >0$, we can find a solution $y$ in the range of $y\in[0,1]$. For instance, using the relation $( 2 \bar{n}_{e} + 1 ) \sinh ( 2 r_{e} ) > ( 2 \bar{n}_{i} + 1 ) \sinh ( 2 r_{i} )$ as clearly seen from Fig.~\ref{fig:1-likelihood-estimation} and Eq.~\eqref{eq:condition-for-equivalence}, we readily derive the squeezing condition $r_{r}$ to have a legitimate solution to Eq.~\eqref{eq:condition-for-equivalence} as
	\begin{equation}
		\sinh ( 2 r_{r} ) > \frac{2 \bar{n}_{e} + 1}{2 \bar{n}_{r} + 1} \sinh ( 2 r_{e} ).
	\end{equation}
The value of $r_{r}$ can be made arbitrarily small by increasing $n_{r}$, but cannot be zero. This points out that although our coarse-graining model is isotropic in the sense that the coarse-graining applies equally to each quadrature in phase-space, it is equivalent only to a phase-sensitive (squeezed) reservoir. Moreover, the effect of coarse-graining is state dependent whereas a Gaussian reservoir affects input states all equally. We may thus say that the quantum-to-classical transitions due to decoherence program and the coarse-grained measurement, respectively, entail unequal features in general.

	\begin{figure}
		\includegraphics[width=\linewidth]{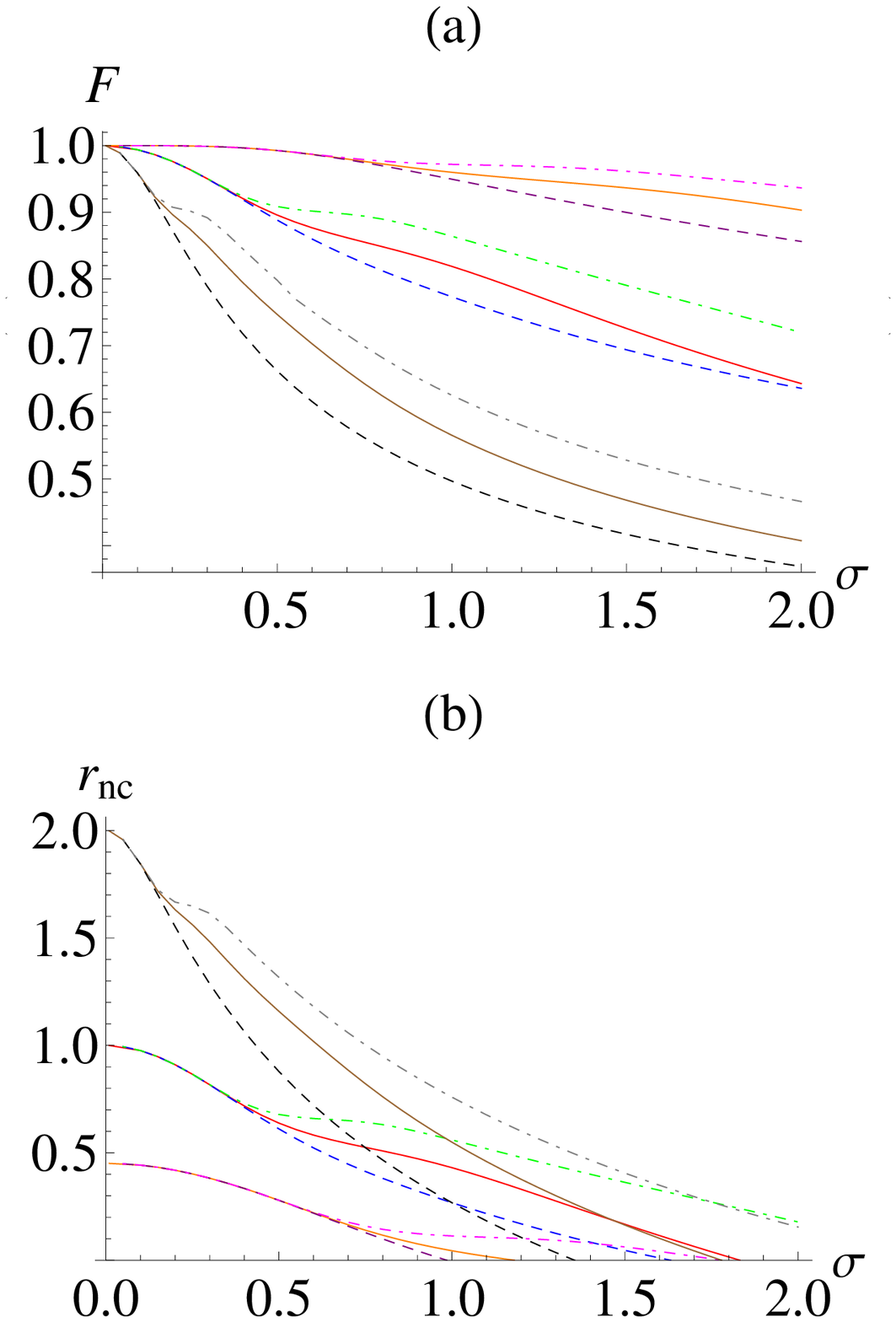}
		\caption{(Color online) (a) Fidelity $F$ between an input state and its reconstructed state and (b) nonclassical squeezing $r_{\mathrm{nc}}$ of the reconstructed state as functions of the coarse-graining size $\sigma$, for the input squeezed thermal states with $( \bar{n}, r ) = ( 0, 1 )$ [green dot-dashed line, red solid and blue dashed lines, the second curves from the top for (a) and (b)], $( \bar{n}, r ) = ( 1, 1 )$ [pink dot-dashed line, orange solid and purple dashed lines, the first curves from the top for (a) and the third curves from the top for (b)], and $( \bar{n}, r ) = ( 0, 2 )$ [gray dot-dashed line, brown solid and black dashed lines, the third curves from the top for (a) and the first curves from the bottom for (b)]. Solid curves represent the case of the MLE method, dot-dashed (dashed) curves the direct reconstruction method with (without) information on the input phase, respectively. For the plots of dashed curves, each point represents an averaged value over the whole range of the input squeezing angles. See main text.}
		\label{fig:1-performance}
	\end{figure}

\subsection{Fidelity and nonclassicality}
From now on, we compare the performance of two estimation methods, the direct reconstruction of the covariance matrix and the MLE, by investigating the fidelity between an input state and its estimated state and the nonclassicality of the estimated state. The fidelity between two single-mode Gaussian states with the same means is given by \cite{Marian2012}
	\begin{equation}
		F^{2} = \frac{1}{\sqrt{\Delta + \Lambda} - \sqrt{\Lambda}},
	\end{equation}
where $\Delta = \det ( \boldsymbol{\Gamma}_{1} + \boldsymbol{\Gamma}_{2} )$ and $\Lambda = 4 \det ( \boldsymbol{\Gamma}_{1} + \frac{i}{2} \mathbf{J} ) \det ( \boldsymbol{\Gamma}_{2} + \frac{i}{2} \mathbf{J} )$ with the symplectic matrix
	\begin{equation}
		\mathbf{J} = \begin{pmatrix} 0 & 1 \\ -1 & 0 \end{pmatrix}.
	\end{equation}
A single-mode Gaussian state is nonclassical, i.e. squeezed, when $r > r_{c} \equiv \frac{1}{2} \ln ( 2 \bar{n} + 1 )$ \cite{Marian1993}. This condition can also be related to the entanglement potential ($\mathcal{P}_{\mathrm{ent}}$) of a single-mode Gaussian state,
	\begin{equation} \label{eq:entanglement-potential}
		\mathcal{P}_{\mathrm{ent}} = \max [ 0, \frac{r - r_{c}}{\ln 2} ],
	\end{equation}
the amount of two-mode entanglement that can be produced by injecting the given state into one mode of 50:50 beam-splitter \cite{Asboth2005}.

In Fig.~\ref{fig:1-performance}, we plot the fidelity $F$ and the nonclassical squeezing $r_{\mathrm{nc}} \equiv \mathcal{P}_{\mathrm{ent}} \ln 2$ of the estimated squeezed thermal state as a function of the coarse-graining size $\sigma$, for the input squeezed thermal states with the parameters $( \bar{n}, r ) = ( 0, 1 )$, $( \bar{n}, r ) = ( 1, 1 )$, and $( \bar{n}, r ) = ( 0, 2 )$. Generally, the larger the coarse-graining size, the worse the fidelity and the nonclassicality. Furthermore, the decrease rate of fidelity and nonclassicality with respect to $\sigma$ becomes larger with increasing input squeezing. 

For the case of direct reconstruction, we also see that the information on the reference frame (phase $\phi_i$ of the input state) plays a crucial role in characterizing the given state. If we have access to the phase information, we may adjust our measurement settings in the direct reconstruction, in which only three measurement angles are chosen in an interval of $\pi/4$ [Eq. (15)], to optimize the characteristics of the reconstructed state (dot-dashed curves). For a given input angle $\phi_i$, the combination of two variances in Eq. (22), which involves $\sigma$ in a nontrivial way, e.g., error function, leads to non-homogeneous behavior in the variance of the reconstructed state, and subsequently the fidelity (dot-dashed curves), as a function of $\sigma$. On the other hand, if we have no access to such phase information, the state characterization generally becomes worse. The dashed curves in Fig. 4 show the results averaged over the squeezing angle $\phi_i$ of the input state, which show worse results than the dot-dashed curves. In those cases, we see that there also exists a rather counter-intuitive regime that a less nonclassical state is more robust to the increment of the coarse-graining, i.e., the crossover of two dashed curves around $\sigma=1$ for the nonclassicality in Fig. 4 (b). The input squeezed thermal state with the parameters $( \bar{n}, r ) = ( 0, 1 )$ retains nonclassical squeezing even when the input state with a larger squeezing, i.e., $( \bar{n}, r ) = ( 0, 2 )$ loses its nonclassicality. It may be also an evidence that this coarse-graining model is not compatible with the decoherence program, where such a crossover does not occur. This feature shows that the coarse-graining on the homodyne measurement has a strong state dependence.

In comparison, the MLE method using a full set of homodyne measurements (solid curves) shows performance at the intermediate level between the direct reconstruction with (dot-dashed line) and without (dashed line) information on the input phase, for both fidelity and nonclassicality. In particular, as the coarse-graining size $\sigma$ becomes rather large, the performance of the MLE is significantly better than that of the direct reconstruction without information on the input phase. Thus, to have access to a reference frame is an important issue in practical situations. 
We note, however, that in an experimentally achievable regime with current technology ($\sigma = 0.1$) \cite{Lvovsky2001, Ourjoumtsev2006, Huisman2009, Cooper2013}, two compared methods may not have a significant difference in their performances. Moreover, in this regime, two methods can detect almost all nonclassical Gaussian states except those with a small nonclassical squeezing $0 < r_{\mathrm{nc}} < 0.0033$. Therefore, it seems practically desirable to adopt the direct reconstruction method, rather than the MLE, as the former requires a fewer number of homodyne measurements, i.e. only three quadrature amplitudes in Eq.~\eqref{eq:single-mode-direct-reconstruction}.

\section{two-mode Gaussian state estimation}
In this section, we extend our study to two-mode Gaussian states reconstructed from the coarse-grained homodyne data. We again investigate the fidelity and the nonclassicality, now entanglement, of the output two-mode state by two coarse-grained methods, the direct reconstruction of the covariance matrix and the MLE. We consider as our input state a two-mode squeezed thermal state (TMST) in the form
	\begin{equation}
		\rho_{2} = \hat{S}_{12} ( r, \phi ) [ \rho_{th} ( \bar{n}_{1} ) \otimes \rho_{th} ( \bar{n}_{2} ) ] \hat{S}_{12}^{\dag} ( r, \phi ),
	\end{equation}
where $S_{12} ( r, \phi ) = \exp [ - r \{ \exp ( i \phi ) \hat{a}_{1}^{\dag} \hat{a}_{2}^{\dag} - \exp ( - i \phi ) \hat{a}_{1} \hat{a}_{2} \} ]$ is the two-mode squeezing operator with the squeezing strength $r$ and the squeezing angle $\phi$. Its covariance matrix is given by
	\begin{equation}
		\boldsymbol{\Gamma} =
			\begin{pmatrix}
				a & 0 & \Re[c] & \Im[c] \\
				0 & a & \Im[c] & - \Re[c] \\
				\Re[c] & \Im[c] & b & 0 \\
				\Im[c] & - \Re[c] & 0 & b
			\end{pmatrix},
	\end{equation}
where
	\begin{align} \label{eq:2CM}
		a & = \bar{n}_{1} \cosh^{2} r + \bar{n}_{2} \sinh^{2} r + \frac{1}{2} \cosh 2r, \nonumber \\
		b & = \bar{n}_{1} \sinh^{2} r + \bar{n}_{2} \cosh^{2} r + \frac{1}{2} \cosh 2r, \nonumber \\
		c & = - \frac{1}{2} ( \bar{n}_{1} + \bar{n}_{2} + 1 ) \exp ( i \phi ) \sinh 2r.
	\end{align}
The characteristic function of TMST is given by
	\begin{align}
		C_{2} ( \lambda_{1}, \lambda_{2} ) & = \exp ( - a | \lambda_{1} |^{2} - b | \lambda_{2} |^{2} - 2 c^{\prime} | \lambda_{1} | | \lambda_{2} | ),
	\end{align}
where $c^{\prime} = |c| \cos ( \varphi_{1} + \varphi_{2} - \phi )$ and $\varphi_{i} = - i \ln ( \lambda_{i} / | \lambda_{i} | )$ ($i \in \{ 1, 2 \}$). From this, we obtain the homodyne distribution as
	\begin{align}
		& P ( x_{1,\varphi_{1}}, x_{2,\varphi_{2}} ) \nonumber \\
		& = \frac{1}{\pi \sqrt{ab - |c|^{2}}} \exp \bigg[ - \frac{b x_{1}^{2} + a x_{2}^{2} + 2 c^{\prime} x_{1} x_{2}}{ab - |c|^{2}} \bigg].
	\end{align}

Similar to the single-mode case, the covariance matrix of a two-mode state can also be constructed by measuring three different quadratures for each mode. The local matrix elements are just the same as in Eq.~\eqref{eq:single-mode-direct-reconstruction}, and the correlation elements are given by
	\begin{align}
		\Gamma_{13} & = \Gamma_{31} = 2 \langle \hat{X}_{1, 0} \hat{X}_{2, 0} \rangle - 2 \langle \hat{X}_{1, 0} \rangle \langle \hat{X}_{2, 0} \rangle, \nonumber \\
		\Gamma_{14} & = \Gamma_{41} = 2 \langle \hat{X}_{1, 0} \hat{X}_{2, \pi / 2} \rangle - 2 \langle \hat{X}_{1, 0} \rangle \langle \hat{X}_{2, \pi / 2} \rangle, \nonumber \\
		\Gamma_{23} & = \Gamma_{32} = 2 \langle \hat{X}_{1, \pi / 2} \hat{X}_{2, 0} \rangle - 2 \langle \hat{X}_{1, \pi / 2} \rangle \langle \hat{X}_{2, 0} \rangle, \nonumber \\
		\Gamma_{24} & = \Gamma_{42} = 2 \langle \hat{X}_{1, \pi / 2} \hat{X}_{2, \pi / 2} \rangle - 2 \langle \hat{X}_{1, \pi / 2} \rangle \langle \hat{X}_{2, \pi / 2} \rangle.
	\end{align}
Using these matrix elements and the relations in Eq.~\eqref{eq:2CM}, we can determine the output two-mode state with the parameters
	\begin{align}
		\bar{n}_{i} & = \frac{(-1)^{i+1} ( a - b ) - 1 + \sqrt{\gamma^{\prime}}}{2}, \nonumber \\
		r & = \frac{1}{2} \mathrm{arcsinh} \Bigg( \frac{2 |c|}{\sqrt{\gamma^{\prime}}} \Bigg), \nonumber \\
		\phi & = \begin{cases} \arctan \Bigg( \dfrac{\Im[c]}{\Re[c]} \Bigg) & \textrm{for} \quad \Re[c] \geq 0, \\
		\pi - \arctan \Bigg( \dfrac{\Im[c]}{\Re[c]} \Bigg) & \textrm{for} \quad \Re[c] < 0, \end{cases}
	\end{align}
with $i \in \{ 1, 2\}$ and $\gamma^{\prime} = ( a + b )^{2} - 4 |c|^{2}$.

The fidelity between two two-mode Gaussian states with same means is given by \cite{Marian2012}
	\begin{equation}
		F^{2} = \frac{1}{\sqrt{\Sigma} + \sqrt{\Lambda} - \sqrt{( \sqrt{\Sigma} + \sqrt{\Lambda} )^{2} - \Delta}},
	\end{equation}
where $\Delta = \det ( \boldsymbol{\Gamma}_{1} + \boldsymbol{\Gamma}_{2} )$, $\Lambda = 16 \det ( \boldsymbol{\Gamma}_{1} + \frac{i}{2} \mathbf{J} ) \det ( \boldsymbol{\Gamma}_{2} + \frac{i}{2} \mathbf{J} )$ and $\Sigma = 16 \det [ ( \mathbf{J} \boldsymbol{\Gamma}_{1} ) ( \mathbf{J} \boldsymbol{\Gamma}_{2} ) - \frac{1}{4} \mathds{1}_{4} ]$ with
	\begin{equation}
		\mathbf{J} = \bigoplus_{i = 1}^{2} \mathbf{J}_{i}, \quad \mathbf{J}_{i} = \begin{pmatrix} 0 & 1 \\ -1 & 0 \end{pmatrix} \quad ( i = 1, 2 ).
	\end{equation}

On the other hand, the entanglement of a two-mode Gaussian state can be measured by the logarithmic negativity \cite{Vidal2002} as
	\begin{equation}
		E_{\mathcal{N}} = \max [ 0, - \log_{2} ( 2 \tilde{\nu}_{-} ) ],
	\end{equation}
where the smaller symplectic eigenvalue $\tilde{\nu}_{-}$ of the partially transposed state is given by
	\begin{equation}
		2 \tilde{\nu}_{-}^{2} = f - \sqrt{f^{2} - 4 g^{2}},
	\end{equation}
with $f = a^{2} + b^{2} + 2 |c|^{2}$ and $g = ab - |c|^{2}$ for the states in our consideration. 

	\begin{figure}
		\includegraphics[width=\linewidth]{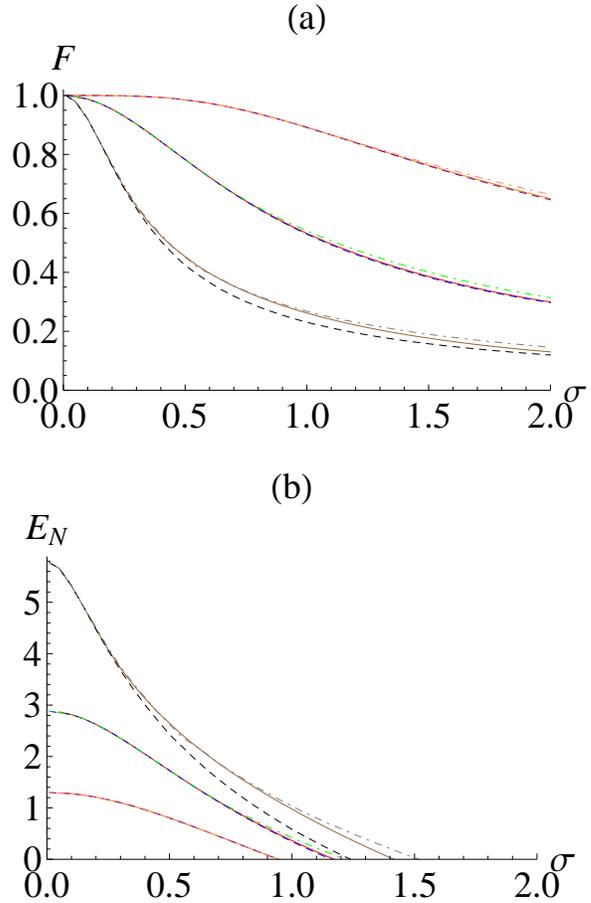}
		\caption{(Color online) (a) Fidelity $F$ between an input state and its reconstructed state and (b) logarithmic negativity $E_{\mathcal{N}}$ of the reconstructed state as functions of the coarse-graining size $\sigma$, for the input squeezed thermal states with $( \bar{n}, r ) = ( 0, 1 )$ [green dot-dashed line, red solid and blue dashed lines, the second curves from the top for (a) and (b)], $( \bar{n}, r ) = ( 1, 1 )$ [pink dot-dashed line, orange solid and purple dashed lines, the first curves from the top for (a) and the third curves from the top for (b)], and $( \bar{n}, r ) = ( 0, 2 )$ [gray dot-dashed line, brown solid and black dashed lines, the third curves from the top for (a) and the first curves from the bottom for (b)]. For simplicity, we assume that the thermal photon number of two modes are the same, $\bar{n}_{1} = \bar{n}_{2} = \bar{n}$. Solid curves represent the case of the MLE method, dot-dashed (dashed) curves the direct reconstruction method with (without) information on the input phase, respectively. For the plots of dashed curves, each point represents an averaged value over the whole range of the input squeezing angles.}
		\label{fig:2-performance}
	\end{figure}

In Fig.~\ref{fig:2-performance}, we plot the fidelity $F$ and the logarithmic negativity $E_{\mathcal{N}}$ of the estimated two-mode squeezed thermal state  as a function of the coarse-graining size $\sigma$, for the input two-mode squeezed thermal states with $( \bar{n}, r ) = ( 0, 1 )$, $( \bar{n}, r ) = ( 1, 1 )$, and $( \bar{n}, r ) = ( 0, 2 )$. For simplicity, we have assumed that the thermal photon number of two modes are the same, $\bar{n}_{1} = \bar{n}_{2} = \bar{n}$. 
These plots show a tendency similar to the plots for the single-mode case in Fig.~\ref{fig:1-performance}. 
The fidelity and the logarithmic negativity decrease with the coarse-graining size $\sigma$, and the degrading rate is larger for a more nonclassical (entangled) initial state. However, each scheme shows a different performance for the characterization of output states. 

The direct reconstruction method without access to the information on the input phase of two-mode squeezing (dashed curves) can generally yield a worse output than that with the information (dot-dashed-curves). For the two-mode squeezed thermal states, the local homodyne distribution for each mode is isotropic as it has no bearing on the phase of two-mode squeezing, so that the estimated mean photon numbers are invariant even when the phase information is not available. Only correlation parts vary under the rotation of the reference frame. As can be seen from Fig.~\ref{fig:2-performance}, the difference in the performance between the two methods is thereby relatively less than that in Fig.~\ref{fig:1-performance}. In addition, the MLE method employing a full set of homodyne measurements again shows performance at the intermediate level. However, the distinctions are not very prominent, and in particular, those three methods yield almost the same results with the currently accessible coarse-graining ($\sigma = 0.1$) \cite{Lvovsky2001, Ourjoumtsev2006, Huisman2009, Cooper2013}. On the other hand, one can readily see that if an asymmetric input state with local squeezings is considered, the availability of the phase information can affect the results more significantly than here.


\section{summary and discussion}
In this paper, we investigated the reconstruction of a quantum state by a coarse-grained homodyne measurement. Employing both the direct reconstruction method of the covariance matrix and the MLE method, we examined single-mode and two-mode Gaussian states to see how those states undergo quantum-to-classical transition. The reconstruction method has been compared to the decoherence model typically employed to account for the quantum-to-classical transition. In particular, as our coarse-graining models produce a Gaussian output state from a Gaussian input state, those models have been compared to the decoherence by a Gaussian reservoir, i.e., thermal squeezed reservoir. We have clearly shown that the coarse-graining model is not compatible with the decoherence model in addressing the state evolution and that the effects (added noise) of coarse-grained reconstruction are particularly state dependent in contrast to the decoherence program. Even though the coarse-graining applies equally to all quadrature amplitudes, i.e., isotropic in phase-space, it turns out that its effect on the state can be made equivalent only by a phase-sensitive reservoir with nonzero squeezing.

Furthermore, we also compared the performance between the direct reconstruction and the MLE in terms of the fidelity and the nonclassicality of the output states. In general, the direct reconstruction method employing homodyne measurement of only three quadratures, therefore practically less demanding, can yield a better output than the MLE method employing a full set of homodyne measurements. However, this is possible only when one has access to the information on the phase of the input state. If the phase information is not available, the MLE method yields better results than the direct reconstruction. In a practical regime of, e.g., $\sigma=0.1$, all those methods yield almost identical results.

As a concluding remark, the reconstruction under a coarse-grained homodyne measurement generally yields a non-Gaussian distribution, i.e., the piecewise flat distribution in Fig. 1. Therefore, even though we know that the input state is a Gaussian state, it will be interesting to study how the characteristics of the reconstructed state can be modified if the MLE method is applied with reference to a set of non-Gaussian states. That is, we take the estimated states to be non-Gaussian and investigate the fidelity and the nonclassicality of the output states, which will be left for future study together with the case of non-Gaussian input states.


\section*{Acknowledgments}
This work is supported by the NPRP Grant No. 4-346-1-061 from the Qatar National Research Fund.

\bibliographystyle{apsrev}

\end{document}